\documentclass[aps,pra,showpacs,twocolumn,amsmath,amssymb,superscriptaddress]{revtex4-1}

\usepackage{graphicx}
\usepackage{subfigure}
\usepackage{mathrsfs}
\usepackage{threeparttable}
\usepackage{dcolumn}
\usepackage{multirow} 
\usepackage{bm}
\usepackage{amsthm}
\usepackage{amsmath}
\usepackage{color}
\usepackage{amsmath}

\theoremstyle{remark}

\newcommand{\la}{\langle}
\newcommand{\ra}{\rangle}
\newcommand{\ket}[1]{| {#1} \ra}
\newcommand{\bra}[1]{\la {#1} |}

\newcommand{\brac}[1]{\lbrace #1\rbrace}

\def\X{\mathcal{X}}
\def\S{\mathcal{S}}
\begin{document}

\preprint{APS/123-QED}

\newcommand{\rmnum}[1]{\romannumeral #1}

\newcommand{\Rmnum}[1]{\expandafter\@slowromancap\romannumeral #1@}

\title{Self-tallying Quantum Anonymous Voting}

\author{Qingle Wang}
\affiliation{State Key Laboratory of Networking and Switching Technology\\
	Beijing University of Posts and Telecommunications, Beijing, 100876, China}
\affiliation{Hearne Institute for Theoretical Physics and
	Department of Physics \& Astronomy\\
	Louisiana State University, Baton Rouge, LA-70820\\}

\author{Chaohua Yu}
\affiliation{State Key Laboratory of Networking and Switching Technology\\
	Beijing University of Posts and Telecommunications, Beijing, 100876, China}

\author{Fei Gao}
\email{gaof@bupt.edu.cn}
\affiliation{State Key Laboratory of Networking and Switching Technology\\
	Beijing University of Posts and Telecommunications, Beijing, 100876, China}

\author{Haoyu Qi}
\affiliation{Hearne Institute for Theoretical Physics and
	Department of Physics \& Astronomy\\
	Louisiana State University, Baton Rouge, LA-70820\\}

\author{Qiaoyan Wen}
\affiliation{State Key Laboratory of Networking and Switching Technology\\
	Beijing University of Posts and Telecommunications, Beijing, 100876, China}

\begin{abstract}
Anonymous voting is a voting method of hiding the link between a vote and a voter, the context of which ranges from governmental elections to decision making in small groups like councils or companies. In this paper, we propose a quantum anonymous voting protocol assisted by two kinds of entangled quantum states. Particularly, we provide a mechanism of opening and permuting the ordered votes of all the voters in an anonymous manner; any party, who is interested in the voting results,  can acquire a permutation copy, and then obtains the voting result through simple calculation. Unlike all previous quantum works on anonymous voting, our quantum anonymous protocol firstly possesses the properties of privacy, self-tallying, non-reusability, verifiability and fairness at the same time. Besides, we demonstrate that the entanglement of the novel quantum states used in our protocol makes the attack from outside eavesdropper and inside dishonest voters impossible. We also generalize our protocol to execute  tasks of anonymous multi-party computation, such as anonymous broadcast and anonymous ranking.

\end{abstract}


\maketitle

\section{\label{sec:level1}Introduction}
\label{sec:1}
Science of cryptography studies how to prevent valuable information from being leaked to unauthorized parties. In practice, most cryptographic protocols are designed to protect messages from being eavesdropped by an adversary when they are sent from one party to another. However, in some situations, to keep the identity of message senders private is just as important as to keep the messages secret. One example is anonymous voting, in which each voter votes for one of candidates anonymously. Therefore, no one but himself or herself could know which candidate he or she votes. The contexts of voting range from governmental elections to decision making in rather
small groups like councils or companies. To be reliable and useful in practice, voting protocols should have some desirable properties (see \cite{AC} for more details) like privacy, non-resusability, verifiability, fairness and eligibility as follows.\\
\indent (1) \emph{Privacy}. Only the individual voter knows how he or she votes.\\
\indent (2) \emph{Non-reusablity}. Each voter can vote only once and cannot change the vote of someone else.\\
\indent (3) \emph{Verifiability}. Each voter can verify whether his or her vote has been counted properly, but cannot prove to anyone else how he or she is voting.\\
\indent (4) \emph{Fairness}. Nobody can obtain a partial vote tally before the end of the protocol.\\
\indent (5) \emph{Eligibility}. Only eligible voters can vote.

In the past decades, a number of voting protocols pursuing above properties have been proposed. The first voting protocol that guarantees voting privacy was proposed
by Chaum in 1981 \cite{C}. Since then various voting protocols based on some cryptographic primitives, such as homomorphic encryption and blind signature, were proposed; a selection of protocols are reviewed in \cite{VSC}.
Most of them adopt public-key cryptographic primitives like large integer factorization and discrete logarithm. However, the found of quantum algorithm makes them not uncrackable anymore \cite{Shor,Grover}. To battle with the power of quantum computer, quantum cryptography was born to encrypt information based upon principle of quantum mechanics. Surprisely, some of these fundamental principles like no-cloning theorem and the observer effect could guarantee unconditional security. Since the first quantum key distribution protocol was proposed in 1984 by Bennett and Brassard \cite{BB84}, a variety of quantum cryptographic protocols have been proposed, including those for key distribution \cite{LC}, secret sharing \cite{HBB,CGL}, coin flipping \cite{ATVY,SR}, private query \cite{JSGBBWZ,GLHW,LGH,WWG}, and so on.

In recent years, researchers have investigated how to use quantum mechanics to preserve the anonymity of senders and receivers in communication tasks. The first quantum protocol to anonymously broadcast classical bits and qubits was proposed by Christandl and Wehner \cite{CW}. Subsequently, much attention has been paid to perform anonymous voting by using quantum principle. Since a quantum anonymous voting protocol \cite{VSC} was presented by Vaccaro et al. in 2007, several quantum anonymous voting protocols \cite{HZBB,DPT,BBHZ} based on entangled states have been put forward. Afterwards, Horoshko and Kilin \cite{HK} gave a quantum anonymous voting protocol which simply utilized single-particle qubit states to vote and Bell states to check the anonymity. More recently, a series of quantum anonymous voting protocols based on continuous variables were proposed in \cite{JHNXZ}. However, these protocols are function-limited from two aspects: (1) most of them only consider two candidates; (2) most of them are designed to achieve only the property of privacy and the other properties are rarely pursed. In special, the property of self-tallying was proposed in classical voting protocol by Kiayias and Yung \cite{KY}, who is interested in the voting result, can tally votes by himself or herself. The functionality of self-tallying avoids the introducing a third party thus reducing the potential risk of security. As far as we know all previous quantum anonymous voting protocols do not have this property, which needs at least one third party to tally votes, and most of them neglect the cheating of third party. e.g., the third party tampers with the voting results.

Is there a quantum voting protocol overcome the above limitations and satisfied all these favorable properties. Actually, it is an interesting question. Here in this paper, we propose the first quantum anonymous voting protocol for any candidates which not only meet privacy, non-resusability, verifiability, fairness, but also has another property, self-tallying. With slightly generalization, we show that our protocol can be used for any anonymous multi-party computation task. This paper is structured as follows. In Sect. \ref{sec:2}, we introduce two kinds of entangled quantum states which will be the key resources to our protocol. We present our self-tallying quantum anonymous voting (SQAV) protocol in Sect. \ref{sec:3}. Then we analyze the security of our protocol in Sect. \ref{sec:4}. In Sect. \ref{sec:5}, we generalize our protocol to anonymous multi-party computation and briefly discuss two possible applications. Finally we discuss the self-tallying, non-reusablitity, verifiability and fairness of our protocol in the Discussion. We draw our conclusion in the last section.

\section{Quantum resources of the protocol}
\label{sec:2}

The security of our SQAV protocol relies on the fact that we use two classes of quantum multiparticle entangled states to distribute the ballot boxes and index numbers to each voter. In this section we introduce  these two states and some properties of them, which we will be use in our protocol.\\

{ Consider a system with $m$ levels with computational basis $\brac{\ket{j}_C, j = 0,1,\cdots,m-1}$. The fourier basis $\brac{\ket{j'}_F, j = 0,1,\cdots,m-1}$, which can be obtained by applying fourier operation on computational basis, is defined as
	\begin{eqnarray}
	\ket{j'}_F = \mathcal{F}\ket{j}_C =\frac{1}{\sqrt{m}}\sum_{k=0}^{m-1}\exp(\frac{2\pi ijk}{m})\ket{k}_C~.
	\end{eqnarray}
}

Now we give the first quantum entangled state in our protocol, which has been dexterously applied to implement the tasks of anonymous voting \cite{DPT} and anonymous ranking \cite{QAR}.

 { The $m$ level $n$-particle state $\ket{\X_n}$ is defined as
\begin{eqnarray}
   \ket{\X_n}=\frac{1}{m^{\frac{n-1}{2}}}\sum_{\sum\limits_{k=0}^{n-1}j_k~\text{mod}~ m=0}|j_0\rangle_C|j_1\rangle_C\cdots |j_{n-1}\rangle_C,
\end{eqnarray}
where $\ket{j_k}$ is the state of $j$th particle in the computational state and  $j_k \in \mathbb{Z}_m := \{0,1,\cdots,m-1\}$.}

$\ket{\X_n}$ has an interesting property that it has the form of GHZ state in the fourier basis,
\begin{eqnarray}
\ket{\X_n}= \frac{1}{\sqrt{m}}\sum_{j=0}^{m-1}|j'\rangle_F|j'\rangle_F\cdots |j'\rangle_F~.
\end{eqnarray}
Therefore $\ket{\X_n}$ has two nice properties. (1) When the state is measured in the computational basis, the summation of the outcomes of each particle modulus $m$ is equal to zero. (2) When the state is measured in the fourier basis, the outcomes of each particle are always the same. To take advantage of above two properties to protect the voting process being eavesdropped or attacked, we need to use the following result \cite{QAR}.\\
\newline
{\bf Theorem 1} {\em A $n$-particle $m$-level quantum state is in the form of $|\mathcal{X}_n\rangle$ if and only if both of the following two conditions are true: (1) when each particle is measured in the computational basis, the sum over all the $n$ measurement outcomes  modulo $m$ is equal to zero; (2) when each particle is measured in the fourier basis, the measurement outcomes are all the same.}\newline

The other quantum entangled states we will use in the voting protocol is defined as follows.

 { A $n$-level $n$-particle singlet state $\ket{\S_n}$ is defined as
 \begin{eqnarray}
   |\mathcal{S}_n\rangle \equiv  \frac{1}{\sqrt{n!}}\sum_{S \in  \mathcal{P}_n^n}(-1)^{\tau(S)}\ket{s_0}\ket{s_1}\cdots\ket{s_{n-1}}.
 \end{eqnarray}
 Here $\mathcal{P}_n^n$ is the set of all permutations of $\mathbb{Z}_n: =\{0,1,\cdots,n-1\}$, $S$ is a permutation (or sequence) in the form $S=s_0s_1\cdots s_{n-1}$. $\tau(S)$, named inverse number, is defined as the number of transpositions of pairs of elements of $S$ that must be composed to place the elements in canonical order, $012\cdots n-1$}.

$|\mathcal{S}_n\rangle$ is $n$-lateral rotationally invariant, which means the measurements of all particles are all different in any basis \cite{CA}. In the Appendix.~\ref{app:X} we give a proof of this property. Specifically,
 \begin{eqnarray}
 \ket{\S_n}_C = e^{i\phi}\ket{\S_n}_F~,
 \end{eqnarray}
where $\phi$ is a phase factor. Again this property will be exploited to ensure the security of the voting protocol through following theorem.\\
\newline
{\bf Theorem 2} {\em
	A $n$-particle $n$-level quantum state is in the form of $|S_n\rangle$, if and only if the following is satisfied: whenever the state is measured in the computational basis or the fourier basis, the permutation of the outcomes of $n$ particles $\brac{s_0,s_1,\cdots,s_{n-1}}$ is a random element of the set $\mathcal{P}_n^n$}.

We give a proof of Theorem 2 in Appendix.~\ref{app:S}.

\section{quantum anonymous voting protocol}
\label{sec:3}
\begin{figure}[b]
\includegraphics[scale=0.7]{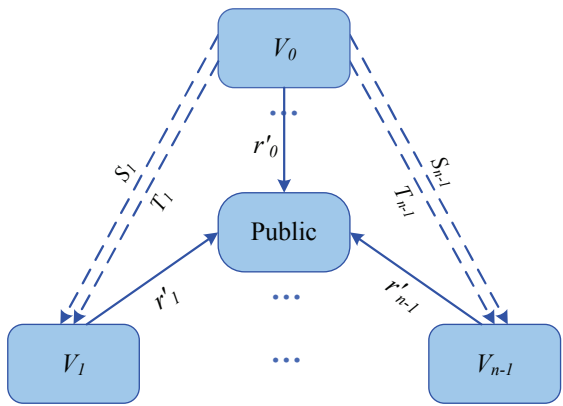}
\caption{\label{fig1} Communications in our protocol. For simplicity, communications in the eavesdropping checks are not considered. The dashed lines represent quantum channels and the solid lines represent classical simultaneous broadcast channels.}
\end{figure}

We first briefly outline our quantum anonymous voting protocol before delving into details. Assume there are $n$ voters labeled as $V_0,V_1,\cdots,V_{n-1}$. Each voter can vote for $m$ candidates labeled by integer $0,1,\cdots,m-1$. Our protocol consists of three steps. First, the entangled state $|\mathcal{X}_n\rangle$ is distributed to $n$ voters. Thus each voter gets $n$ secret `ballot boxes', each of which contains a random number, called the \emph{ballot number}. Then the entangled state $|\mathcal{S}_n\rangle$ is distributed to $n$ voters. Thus each voter gets a random number called \emph{index number}, which decides which ballot box each voter will use for their voting. Second, each voter casts a vote to his or her indicated ballot box anonymously and all voters open all ballot boxes at the same time. Finally, a random permutation of all votes is appear and any party, who is interested in the voting result, can obtain a copy of permutation thus having the voting result. The details of our protocol are presented as follows and the communications in our protocol are shown in Fig. \ref{fig1}.

\subsection{Procedure of the protocol}
\noindent{\bf Step 1. Distributing secret ballot boxes.}\\\\
(1.1) {\em Prepare quantum states}. \\
One of $n$ voters is chosen randomly to prepare $n+n\delta_0$ copies of quantum state $|\mathcal{X}_n\rangle$, where $\delta_0$ is the security strength. Without loss of the generality, we assume $V_0$ is appointed as the distributor. The $j$th copy of state $\ket{\X_n}$ lives in the Hilbert space of $n$ particles, $p_{j,0},p_{j,1}, \cdots, p_{j,{n-1}}$. Therefore we have a {\em particle matrix}, $p_{j,k}$ with $0\leq j \leq n+n{\delta_0-1}, 0\leq k \leq n-1$.\\\\
(1.2) {\em Distribute to each voter}
\\The distributor $V_0$ sends each column of the particle matrix, $S_k = \brac{p_{0,k},p_{1,k},\cdots,p_{n+n{\delta_0-1},k}}$, to each voter $V_k$ ($V_0$ keeps $S_0$).\\\\
(1.3) {\em Security test}\\
After each voter has received his or her particle sequence, each voter as the checker performs the security check processes to ensure the state distributed is intact. Start from voter $V_0$ (the order does not matter), he or she randomly picks out $\delta_0$ particles as the test particles,
\begin{eqnarray}
\vec{p}^{0}_{\text{test}} = p_{i_0,0}p_{i_1,0}...p_{i_{\delta_0-1},0}~.
\end{eqnarray}
$V_0$ also needs to choose randomly from computational basis or fourier basis with uniform distribution for each test state, in which he or she will measure his or her test particles with chosen basis. Then he or she publishes the row index of his or her test particles and the measurement basis he or she chosen to do the measurement. After receiving this information, all other voters are required to measure their particles with the same row index,
\begin{eqnarray}
\vec{p}^{k}_{\text{test}} = p_{i_0,k}p_{i_1,k}...p_{i_{\delta_0-1},k},~~ k = 1,2,\cdots,n-1~.
\end{eqnarray}
in the basis picked by the checker $V_0$. In other words, the $i_0$th, $i_1$th, $\cdots$, $i_{\delta_0-1}$th copies of $|\mathcal{X}_n\rangle$ are samples and measured in either the computational basis or fourier basis. Then all voters send their measurement outcomes to the checker $V_0$ in the order designed by $V_0$.
Let's label the results of measuring each test particle as $r_{i_j,k}$. If $V_0$ chooses the computational basis, he or she then needs to check if $\sum\limits_{j=0}^{n-1} r_{i_j,k} ~\text{mod}~ m = 0$. If $V_0$ chooses the fourier basis, he or she needs to verify whether $r_{i_j,0},r_{i_j,1},\cdots,r_{i_j, n-1}$  are all same. If the test is failed, $V_0$ informs the other voters to abort the protocol. If the test is passed, the same test procedure is performed by the next checker. Repeat the same procedure until the test performed by each voter is passed or abort the protocol in some intermediate step. \\\\
(1.4) {\em Generate ballot numbers}\\
 If the security test passes, each voter now has $n$ particles left after discarding all test particles. Each voter then measures his or her $n$ particles in the computational basis. This will generate $n$ ballot numbers for each voter. Ballot numbers of all voters form a  {\em ballot matrix}, $r_{j,k}\in \{0,1,\cdots,m-1\}$. The $k$th colomn are the private ballot numbers for $V_k$. Since the security is passed, each left copy of $\ket{\X_n}$ is intact, according to theorem 1, ballot numbers must satisfy the condition
\begin{eqnarray}
   \sum_{k=0}^{n-1}r_{j,k}~\textmd{mod}~m =0~.
\end{eqnarray}
for $j=0,1,\cdots,n-1$.\\\\
{\bf Step 2. Distributing secret indexes.}\\\\
(2.1) {\em Prepare quantum states}. \\
Similarly to step (1.1), one of $n$ voters is chosen randomly to prepare $1+n\delta_1$ copies of quantum state $|\mathcal{S}_n\rangle$, where $\delta_1$ is the security strength. The $j$th copy of state $\ket{\S_n}$ lives in the Hilbert space of $n$ particles, $t_{j,0},t_{j,1},\cdots,t_{j,{n-1}}$. Therefore we have a {\em particle matrix}, $t_{j,k}$ with $0\leq j \leq n\delta_1, 0\leq k \leq n-1$.\\\\
(2.2) {\em Distribute to each voter}
\\The distributor sends each column of the particle matrix, $T_k = \brac{t_{0,k},t_{1,k},\cdots, t_{n\delta_1,k}}$, to the voter $V_k$ .\\\\
(2.3) {\em Security test}\\
After each voter has received his or her particle sequence, each voter performs the security check processes to ensure the state distributed is intact. Start from voter $V_0$ (the order does not matter), he or she randomly picks out $\delta_1$ particles as the test particles,
\begin{eqnarray}
\vec{t}^{0}_{\text{test}} = t_{i_0,0},t_{i_1,0},\cdots,t_{i_{\delta_1-1},0}~.
\end{eqnarray}
$V_0$ also needs to choose randomly from computational basis or fourier basis with uniform distribution for each test particle, in which he or she will measure his or her test particle with chosen basis. Then he or she publishes the row index of his or her test particles and the corresponding measurement basis he or she chosen to do the measurement. After receiving this information, all other voters are required to measure their particles with the same row index,
\begin{eqnarray}
\vec{t}^{k}_{\text{test}} = t_{i_0,k},t_{i_1,k},\cdots,t_{i_{\delta_1-1},k}~,
\end{eqnarray}
for $k = 0,1,2,\cdots,n-1$ in the basis picked by the checker $V_0$ and send their measurement outcomes to the checker $V_0$ in the order appointed by $V_0$. That is, the $i_0$th, $i_1$th, $\cdots$, $i_{\delta_1-1}$th copies of $|\mathcal{S}_n\rangle$ are measured in either the computational basis or the fourier basis.
Label the results of measuring each test particle as $d_{i_j,k}$. No matter $V_0$ choose the computational basis or the fourier basis, he or she then needs to check if $\brac{d_{i_j,0},d_{i_j,1},\cdots,d_{i_j,{n-1}}}\in \mathcal{P}_n^n $ according to theorem 2. If the test is passed, the same test procedure is performed by the next checker. If the test is failed, $V_0$ informs the other voters to abort the protocol. The same procedure is repeated until the test performed by each voter is passed or the protocol is aborted in some certain intermediate step. \\\\
(2.4) {\em Generate index numbers}\\
If the security test passes and then discards all test particles, each voter now has only one particle left. Each voter then measures his or her particle in the computational basis. This will generate an index number for each voter. Index numbers of all voters form an {\em index array}, $d_{k}\in \{0,1,\cdots,m-1\}$. $d_{k}$ indicates anonymously that $d_k$th ballot box is the box for $V_k$ to cast vote. Since the security has tested, the only left copy of $\ket{\S_n}$ is intact according to theorem 2. Here $d_0,d_1,\cdots,d_{n-1} \in \mathcal{P}_n^n$.
\\\\
{\bf Step 3. Vote casting.}\\\\
(3.1) {\em Vote casting}\\
After steps 1 and 2, each voter $V_k$ has $n$ ballot numbers, $r_{0,k},r_{1,k},\cdots,r_{n-1,k}$, and one index number, $d_k$. Now voter $V_k$  votes to the candidate $v_k \in \{0,1,\cdots,m-1\}$,  by adding $v_k$ to $r_{d_k,k}$. He or she then renews ballot numbers $r_{jk}'=(r_{0,k}',r_{1,k}',\cdots,r_{n-1,k}')$, in which
\begin{eqnarray}
\label{EQ:6}
r_{j,k}' =
\begin{cases}
r_{j,k} + v_k ~\text{mod m}~ &~\text{if}~j = d_k~,\\
r_{j,k}&~\text{if}~j\neq d_k~.
\end{cases}
\end{eqnarray}
All voters publish all the updated ballot numbers through simulation broadcast channels \cite{BT,HM}.
At last we have a {\em vote matrix}, $r_{j,k}'$, which is available for every party at the same time.\\\\
(3.2) {\em Self-tallying}\\\\
With the vote matrix, each party, who is interested in the voting result, can count the votes for each candidate. They take the summation of each row,
\begin{eqnarray}\label{EQ:tally}
R_j&=&\sum_{k=0}^{n-1}r_{j,k}^{'}~\textmd{mod}~m~,\\
&=& \sum_{k=0}^{n-1}r_{j,k} + v_{k_0}~.
\end{eqnarray}
Here $d_{k_0} = j$. Therefore $\brac{R_0,R_1,\cdots,R_{n-1}}$ is a permutation of the votes $\brac{v_0,v_1,\cdots,v_{n-1}}$. The number of votes each candidate got is given by
\begin{eqnarray}
N_i = \sum_{R_j=i}1~,
\end{eqnarray}
for $i = 0,1,\cdots,m-1.$\\\\
(3.3) {\em Security check}\\\\
Each voter $V_k$ needs to verify that $R_{d_k} = v_k$. If the answer is yes, it indicates that his or her vote is counted correctly; otherwise the protocol is aborted since the voting step is compromised.

\subsection{Example}
To illustrate the protocol, we give a simple example (see Table. \ref{Tab:1}) with $n=4$ voters and $m=3$ candidates. For simplicity, we assume there is no eavesdrop or attack happened. Thus we ignore the security tests (steps (1.3), (2.3) and (3.3)). After executing step 1, suppose ballot matrix held by 4 voters are
\begin{eqnarray}
r_{j,k}=
\begin{pmatrix}
0 & 1 & 2 & 0\\
2 & 2 & 1 & 1\\
1 & 0 & 2 & 0\\
0 & 1 & 1 & 1
\end{pmatrix}~.
\end{eqnarray}
 After step 2, assume the index numbers are
 \begin{eqnarray}
 d_k = (1,0,3,2)~.
 \end{eqnarray}
Then in step 3,  assume the four voters $V_0$, $V_1$, $V_2$ and $V_3$ cast votes
\begin{eqnarray}
v_k = (1,2,1,0).
\end{eqnarray}
The voting and self-tallying process are described in Table. ~\ref{Tab:1}. The final published results are
\begin{eqnarray}
R_j = (2,1,0,1)
\end{eqnarray}
which is indeed a permutation of the votes $v_k$ as we expected.

\begin{table}[h]
\centering{Example of QAVP}\\
\begin{ruledtabular}
\begin{tabular}{llllll}
          &$V_0$ &$V_1$ &$V_2$ &$V_3$ &$R_j$ \\ \hline
$r_{0,k}'$ &0     &1+2   &2     &0     &2     \\
$r_{1,k}'$ &2+1   &2     &1     &1     &1     \\
$r_{2,k}'$ &1     &0     &2     &0+0   &0     \\
$r_{3,k}'$ &0     &1     &1+1   &1     &1     \\
\end{tabular}
\end{ruledtabular}
\caption{\label{Tab:1}%
	A simple example of QAVP with $n=4$ and $m=3$. Each voter adds his or her votes to the ballot assigned by his or her index number. The tallying results are calculated according to Eq. (\ref{EQ:tally}).}
\end{table}
\section{Privacy analysis}
\label{sec:4}
Privacy is the primary property of a SQAV protocol. In this section, we focus on discussing the privacy of our SQAV and other properties will be given in section VI. Generally, the top priority is to protect the privacy of each voter, namely no outsider or voters should know which vote is cast by whom, except the one by himself or herself. In our SQAV, the attacker could be an outside eavesdropper, one dishonest voters  \cite{GQWF,LGGZ} or the adversary which concludes some dishonest voters. If an attacker successfully eavesdrops the ballot random numbers or index number of the voter $V_k$ without being detected, he or she can easily know which candidate $V_k$ votes for. Therefore, preserving privacy in our SQAV requires preventing ballot numbers and index numbers from being eavesdropped.
The security tests in steps (1.3) and (2.3) are designed to protect the ballot matrix, index array and the voting process from being compromised.
\subsection{Outside eavesdropper}
For outside eavesdropper, Eve could intercept the $S_k$ or $T_k$ during step (1.2) or (2.2). Let's consider that Eve intercepts arbitrary $x$ particles she would like to in $S_k$. If $x < n$, then there is a chance that all $x$ particles are happen to be among the $n$ particles which are not included in the tests. Actually the probability of this happening is
\begin{eqnarray}\nonumber
P_e &=& {{n}\choose{x}}\big/{{n+n\delta_0}\choose{x}}\\\nonumber
&=& \frac{n!}{(n-x)!}\frac{(n+n\delta_0-x)!}{(n+n\delta_0)!}\\
&=& \prod_{k=n}^{n-x+1}\frac{k}{k+n\delta_0}\\
&\sim & \mathcal{O}((\frac{1}{\delta_0})^x)~,
\end{eqnarray}
which is approaching to zero if we make the security strength $\delta_0$ large enough.
Actually the more particles Eve intercepts, the faster the probability that she could pass the security check goes to zero. Similarly we could argue that the probability of Eve intercepting and modifying $T_k$ in Step 2 without being found is negligible. Therefore, for large enough $\delta_0$, $\delta_1$, the disturbed particles cannot escape from the security tests in steps (1.3) and (2.3).

Let's consider another scenario. Assume Eve intercepts and modifies $p_{j_0,k}$ in $S_k$ thus the $j_0$th copy of $\ket{\X_n}$ is changed. Suppose that the new state due to Eve's disturbance is $\ket{\phi_e}$. The probability of all security tests in Step (2.3) are passed is
\begin{eqnarray}
P_e = (\frac{1}{2}P_C +\frac{1}{2}P_F)^{n\delta_0}~,
\end{eqnarray}
where
\begin{eqnarray}
P_C &=& \sum_{\sum_k j_k \text{mod}~m=0} ~|\bra{\phi_e}j_0,j_1,\cdots,j_{n-1}\rangle_C|^2~,\\
P_F &=& \sum_{j=0}^{m-1}|\bra{\phi_e}j,j,\cdots,j\rangle_F|^2~.
\end{eqnarray}
Since $\bra{\phi_e}\X_n\rangle\neq 1$ according to theorem 1, $P_C+P_F < 1$. Therefore, for large enough $\delta_0$,
\begin{eqnarray}
P_e\rightarrow 0~.
\end{eqnarray}
 The argument for Eve modifying the index number is similar, based on theorem 2, if $\delta_1$ is large enough Eve cannot pass the security tests. In summary, as long as the security strength $\delta_0, \delta_1$ are large enough, the attack from outside eavesdropper can be prevented.

\subsection{The dishonest voters cannot eavesdrop the ballot numbers without being detected}
In the step 1, to gain the information of ballot numbers of honest voters, the dishonest voters could cooperate to attack the particles during their transmission in step (1.2) and announce the wrong results to avoid being detected by the honest voters in step (1.3). Since $V_0$ is the only voter who prepares and distributes the quantum states, it seems that $V_0$ plays a different role from the other voters. To analyze the possible attacks from dishonest voters in more detail, two cases: (1) $V_0$ is honest and (2) $V_0$ is dishonest, are considered.

For the case (1), without loss of generality, we suppose there are $l$ dishonest voters, $V_{i_0},V_{i_1},\cdots,V_{i_{l-1}}$. The most general attack by the dishonest voters is that they intercept some particles any they would like to during the transmission from $V_0$ to honest voters and then they perform an unitary operation (attack operation) on intercepted particles and an auxiliary system to yield a new state, denoted by $|\Phi\rangle$, of the composite system. To avoid being detected by the honest voters in step (1.3) when they measure their particles in their hands with the fourier basis and the measurement outcomes are required to be the same, $|\Phi\rangle$ should be in the form
\begin{eqnarray}
|\Phi\rangle=\frac{\sum\limits_{j=0}^{m-1}|j'\rangle_{0}|j'\rangle_{{j_0}} \cdots |j'\rangle_{{j_{n-l-2}}} |\phi_j\rangle}{\sqrt{m}},
\end{eqnarray}
where $|\phi_j\rangle$ are the states of the composite system of $l$ particles sent from $V_0$ to the dishonest voters and the auxiliary system (denoted by system $E_0$), and the subscripts $0,j_0,j_1,\cdots,j_{n-l-2}$ represent the particles held by honest voters $V_0,V_{j_0},V_{j_1},\cdots,V_{j_{n-l-2}}$. It can be rewritten in the computational basis as
\begin{eqnarray}
|\Phi\rangle =\sum\limits_{k_{0},k_{j_0},\cdots,k_{j_{n-l-2}}=0}^{m-1} &&\frac{|k_{0}\rangle |k_{j_0}\rangle \cdots|k_{j_{n-l-2}}\rangle}{m^{\frac{n-l+1}{2}}}\nonumber \\
&&\otimes |\varphi_{k_{0}k_{j_0}\cdots k_{j_{n-l-2}}}\rangle,
\end{eqnarray}
where $|\varphi_{k_{0}k_{j_0}\cdots k_{j_{n-l-2}}}\rangle=\sum\limits_{j=0}^{m-1}\exp{(\frac{2\pi ij(k_{0}+k_{j_0}+\cdots+ k_{j_{n-l-2}})}{m})}|\phi_j\rangle$ are the unnormalized state vectors of system $E_0$. The dishonest voters could measure the system $E_0$ and obtain some $|\varphi_{k_{0}k_{j_0}\cdots k_{j_{n-l-2}}}\rangle$ to infer the measurement outcomes ${k_{0}k_{j_0}\cdots k_{j_{n-l-2}}}$ of honest voters in step (1.4). From the form of $|\varphi_{k_{0}k_{j_0}\cdots k_{j_{n-l-2}}}\rangle$, it is easy to see that, for any two different outcomes ${k_{0}k_{j_0}\cdots k_{j_{n-l-2}}}$ and ${k'_{0}k'_{j_0}\cdots k'_{j_{n-l-2}}}$ such that ${k_{0}k_{j_0}\cdots k_{j_{n-l-2}}}={k'_{0}k'_{j_0}\cdots k'_{j_{n-l-2}}} \ \textmd{mod} \ m$ , $|\varphi_{k_{0}k_{j_0}\cdots k_{j_{n-l-2}}}\rangle$ = $|\varphi_{k'_{0}k'_{j_0}\cdots k'_{j_{n-l-2}}}\rangle$. This means that the dishonest voters can only at most know the information about the sum ${k_{0}k_{j_0}\cdots k_{j_{n-l-2}}} \ \textmd{mod} \ m$ by measuring the system $E_0$. However, this attack is trivial in the sense that without any eavesdropping attack the dishonest voter can cooperate to directly infer the sum of measurement outcomes (ballot numbers) of honest voters after executing the step (1.4).

For the case (2) that $V_0$ is dishonest, we assume there are other $l$ dishonest voters  $V_{i_0},V_{i_1},\cdots,V_{i_{l-1}}$. The most general attack for them are similar to the case (1). The only difference could be that the dishonest voters can directly prepare and distribute fake states to the honest voters instead of intercepting the particles. To avoid being detected by honest voters, these states should be of the form similar to Eq. (25) or (26). From the above analysis for case (1), it is not hard to draw the same conclusion as case (1) that, in order to avoid being detected, the dishonest voters can only perform a trivial attack to obtain the sum of ballot numbers of the honest voters.

\subsection{The dishonest voters cannot eavesdrop the index numbers without being detected}

In step 2, to eavesdrop the information of index numbers of honest voters, the dishonest voters could also attack the particles during their transmissions in step (2.2) and announce the wrong results to avoid being detected by the honest voters in step (2.3). Just as analyzing eavesdropping the ballot numbers in the last subsection, we also consider two cases: (1) $V_0$ is honest and (2) $V_0$ is dishonest.

For the case (1), we also assume there are $l$ dishonest voters, $V_{i_0},V_{i_1},\cdots,V_{i_{l-1}}$. The most general attack for them is that, they first intercept some transmitted particles in step (2.2), entangle them with an auxiliary system prepared in advance and then return the operated particles to honest voters. The state of the whole composite system is denoted by $|\Psi\rangle$. To elude detection in step (2.3), it is required that all the measurement outcomes should be distinct when measuring each particle held by honest voter in the fourier basis, and thus $|\Psi\rangle$ should be in the form
\begin{eqnarray}
|\Psi\rangle = \sum_{S \in \mathcal{P}_n^{n-l}}&&\frac{(-1)^{\tau(S)}\mathcal{F}^{\otimes(n-l)}|S\rangle}{\sqrt{|\mathcal{P}_n^{n-l}|}}\otimes|u_{S}\rangle ,
\end{eqnarray}
where $S=s_{0}s_{j_0}\cdots s_{j_{n-l-2}}$, $|u_{S}\rangle$ are the states of composite system (denoted by $E_1$) of $l$ particles sent to the dishonest voters and auxiliary system, $\mathcal{P}_n^{n-l}=\{x_0 x_1\cdots x_{n-l-1}| x_0,x_1,\cdots,x_{n-l-1} \in \mathbb{Z}_n, \forall j \neq k, x_j \neq x_k\}$ is the set of all the $(n-l)$-permutations of $\mathbb{Z}_n$ and $|\mathcal{P}_n^{n-l}|=\frac{n!}{l!}$ is its size. $\mathcal{P}_n^{n-l}$ can be divided into ${{n}\choose{n-l}}=\frac{n!}{(n-l)!l!}$ subsets, each one corresponding to the set of all the $(n-l)!$ permutations of a $(n-l)$-combination of $\mathbb{Z}_n$. In addition, any two states $|u_{S_0}\rangle$ and $|u_{S_1}\rangle$ such that $S_0 \in \mathcal{P}_n^{n-l,w_0}$, $S_1 \in \mathcal{P}_n^{n-l,w_1}$ and $w_0 \neq w_1$ should be orthogonal to each other, i.e., $\langle u_{S_0}|u_{S_1}\rangle=0$; if not, the dishonest voters cannot deterministically know subset $\mathcal{P}_n^{n-l,w}$ in which the honest voters' measurement outcomes are, and thus cannot announce the correct measurement outcomes to avoid being detected. Rewriting $|\Psi\rangle$ in the computational basis, we have
\begin{eqnarray}
\label{EQ:11}
|\Psi\rangle = \frac{n^{-\frac{n-l}{2}}}{\sqrt{|\mathcal{P}_n^{n-l}|}}\sum_{ T \in \mathcal{R}_n^{n-l}}&&|T\rangle \otimes|v_{T}\rangle,
\end{eqnarray}
where $T = t_0 t_{j_0}\cdots t_{j_{n-l-2}}$, $\mathcal{R}_n^{n-l}=\{x_0 x_1\cdots x_{n-l-1}| x_0,x_1,\cdots,x_{n-l-1} \in \mathbb{Z}_n\}$ and
\begin{eqnarray*}
|v_{T}\rangle=\sum_{S \in \mathcal{P}_{n}^{n-l}}(-1)^{\tau(S)}\exp(\frac{2\pi i(s_0t_0+\sum_{k=0}^{n-l-2}s_{j_k}t_{j_k})}{n})|u_{S}\rangle\\
=\sum_{w}\sum_{S \in \mathcal{P}_{n}^{n-l,w}}(-1)^{\tau(S)}\exp(\frac{2\pi i(s_0t_0+\sum_{k=0}^{n-l-2}s_{j_k}t_{j_k})}{n})|u_{S}\rangle
\end{eqnarray*}
are the unnormalized state vectors of system $E_1$. To avoid being detected by the honest voters who measure their particles in the computational basis in the step (2.3) and the measurement outcomes are required to be distinct, two conditions should be satisfied:
(a) in Eq. (\ref{EQ:11}) there is no terms $|v_{T}\rangle$ for $T \notin \mathcal{P}_n^{n-l}$, or equivalently,  for $T \in \mathcal{Q}_n^{n-l}=\{x_0 x_1\cdots x_{n-l-1}| x_0,x_1,\cdots,x_{n-l-1} \in \mathbb{Z}_n, \exists j \neq k, x_j=x_k\}$;
(b) any two states $|v_{T_0}\rangle$ and $|v_{T_1}\rangle$ for $T_0 \in \mathcal{P}_n^{n-l,w_0}$, $T_1 \in \mathcal{P}_n^{n-l,w_1}$ and $w_0 \neq w_1$ should be orthogonal to each other, i.e., $\langle v_{T_0}|v_{T_1}\rangle=0$. Here we focus on analyzing what $|\Psi\rangle$ (in Eq. ($\ref{EQ:11}$)) should be to satisfy the condition (a). Since $\langle u_{S_0}|u_{S_1}\rangle=0$ for $S_0 \in \mathcal{P}_n^{n-l,w_0}$, $S_1 \in \mathcal{P}_n^{n-l,w_1}$ and $w_0 \neq w_1$, the condition (a) is equivalent to the one that  $\sum_{S \in \mathcal{P}_{n}^{n-l,w}}(-1)^{\tau(S)}\exp(\frac{2\pi i(s_0t_0+\sum_{k=0}^{n-l-2}s_{j_k}t_{j_k})}{n})|u_{S}\rangle=0$ for arbitrary $w$ and arbitrary $T \in \mathcal{Q}_n^{n-l}$. To satisfy this condition, for arbitrary $w$, all the $|u_{S}\rangle$ such that $S \in \mathcal{P}_n^{n-l,w}$ should be equal (denoted by $|u_{w}\rangle$), which is implied by the the corollary 1 of Appendix. Thus $|v_{T}\rangle$ can be rewritten as
\begin{eqnarray}
\label{EQ:12}
|v_{T}\rangle &=&\sum_{w}\sum_{S \in \mathcal{P}_{n}^{n-l,w}}(-1)^{\tau(S)}\nonumber \\
&&\exp(\frac{2\pi i(s_0t_0+\sum_{k=0}^{n-l-2}s_{j_k}t_{j_k})}{n})|u_{w}\rangle.
\end{eqnarray}
Once the dishonest voters successfully elude the eavesdropping check process in step (2.3), they could measure the system $E_1$ and get some $|v_{T}\rangle$ to infer the index numbers $T=t_0t_{j_0},\cdots t_{j_{n-l-2}}$ of honest voters in step (2.4). However, from the form of $|v_{T}\rangle$ in Eq. (\ref{EQ:12}), it is easy to verify that for any two sequences $T_0$, $T_1$ which are in the same subset $\mathcal{P}_{n}^{n-l,w}$, $|v_{T_0}\rangle=|v_{T_1}\rangle$. Therefore, the dishonest voters can at most know the information about which subset (i.e., $w$) the honest voters' index numbers are in. However, this general entangle-measure attack is trivial in the sense that without any attack the dishonest can cooperate to obtain this information.

For the case (2) that $V_0$ is dishonest, the general attack performed by them would be the same as the case (1) except that the dishonest voters would prepare and distribute the fake states in the form similar to the Eq. (27) to the honest voters instead of intercepting the particles in step (2.2). According the analysis in case (1), we can conclude that the dishonest voters cannot obtain the index numbers of honest voters without being detected.

\section{Generalize to anonymous multi-party computation}
\label{sec:5}
One important feature of SQAV is to make each vote open without any relation with any voter. Actually it provides a mechanism to implement a class of multi-party tasks. That is, our protocol can be as useful as for voting as long as a multi-party activity requires to broadcast the data of each party anonymously. Therefore we define a more general class of problem, anonymous multi-party computation (AMC) as follows.\\

\noindent{\bf Definition} {\em Anonymous multi-party computation is a task to compute a function of the form $f(y_0^0,\cdots,y_0^{i_0-1},y_1^0,\cdots,y_1^{i_1-1},y_{n-1}^0,\cdots,y_{n-1}^{i_{n-1}-1})$ by $n$ parties. The function $f$ is invariant under the permutation of integer inputs $\brac{y_k^i}$. Each party, say $P_k$, feeds $y_{k}^0,\cdots,y_{k}^{i_{j}-1}$ in the function anonymously and obtains the result unassisted the other person. All the inputs are bounded by $0\leq y_k< m$.}\\

\noindent The protocol for AMC is very similar to the SQAV.

Step 1. $P_0$ prepares $\bar{n}+n\delta_2$ copies of $m$ level $n$-particle state $\ket{\X_{n}}$, where $\bar{n} = \sum_{k=0}^{n-1}i_k$. Then $P_0$ keeps the first $i_0$ columns  $S_0,S_1,\cdots,S_{i_0-1}$ to himself and then distribute $S_{\sum_{t=0}^{k-1}i_{t}-1},\cdots,S_{\sum_{t=0}^{k}i_{t}-1}$ to $P_k$. After distribution, each party $P_k$ repeats the security procedure in step (1.3). If all $n$ tests are passed, each parity $P_k$ measures his or her $\bar{n}$ particles so again there is a ballot column
\begin{eqnarray}
r_{j,k}=
\begin{pmatrix}
r_{0,k}\\
r_{1,k}\\
\vdots\\
r_{\bar{n}-1,k}
\end{pmatrix}~.
\end{eqnarray}

Step 2. $P_0$ prepares $1+n\delta_3$ copies of $\ket{\S_{\bar{n}}}$ and distribute $T_{\sum_{t=0}^{k-1}i_{t}-1},\cdots,T_{\sum_{t=0}^{k}i_{t}-1}$ to $P_k$. Again to protect from attack, each party is required to choose $\delta_3$ copies of $\ket{\S_{\bar{n}}}$ to exam if $\ket{\S_{\bar{n}}}$ is intact. If all tests are passed, each party $P_k$ measures the remaining particles with computation basis and then there are index arrays $d_{0,k},\cdots,d_{i_k,k}$, where  $d_{i,k}\in\brac{0,1,\cdots,\bar{n}-1}$.

Step 3. Finally each party addes each of his or her data to the ballot box decided by the corresponding index number. And we have a {\em data matrix} $r_{j,k}'$. Finally every party could calculate
\begin{eqnarray}
R_j = \sum_{k=0}^{n}r'_{j,k} \mod m~,
\end{eqnarray}
$\brac{R_j}$ is a permutation of all the data $\bigcup_{j=0}^{n
-1}y_j^i$. Therefore all the data are broadcasted anonymously.

Step 4. With holding all data, each party can obtain the result of $$f(y_0^0,...,y_0^{i_0-1},y_1^0,...,y_1^{i_1-1},y_{n-1}^0,\cdots,y_{n-1}^{i_{n-1}-1})$$ through simple calculation by himself or herself.

Actually, AMC is a subclass of secure multi-party computation (SMC) problem, in which a number of parties also jointly compute
a function over their inputs while the inputs are kept private. SMC focuses on function result without publication of the inputs of all parties. To illustrate it, we give a simple example in which three parties want to jointly compute the function $f(y_0,y_1,y_2)=y_0+y_1+y_2$ over their inputs $y_0$, $y_1$ and $y_2$. Supposing $y_0=2, y_1=3, y_2=6$, by SMC, they have the result $f(y_0,y_1,y_2)=11$, but any party can only know the sum of the inputs of the other two parties. However, by AMC, in addition to obtaining the result $f(y_0,y_1,y_2)=11$, every party also get a permutation of the original inputs of the other , for example, $(3,6,2)$ and the index of his or her own input is only known to himself or herself; as a result, $P_0$ knows $(y_1,y_2)=(3,6)$ or $(6,3)$, $P_1$ knows $(y_0,y_2)=(2,6)$ or $(6,3)$, $P_2$ knows $(y_0,y_1)=(2,3)$ or $(3,2)$. However, for some particular tasks, the function result leads to open all inputs. In this sense, there is no difference between AMC and SMC. In the following, we give two examples to explain this.
\subsection{Anonymous broadcast}
The simplest application of AMC is to implement anonymous broadcast (AB). AB channels are primitives of many anonymous communication protocols.

An anonymous $n$-party broadcast task \cite{CW} is to publish the datum $y_k\in\{0,1,\cdots,m-2\}$ held by sender $P_k$ anonymously and all parties obtain $y_k$ at the same time. In this scenario, the protocal are basically the same as of SQAV with $m$ candidate and $n$ voters. If a sender would like to broadcast message $y$, he or she just needs to `vote' for the `candidate' $y$ following the protocol in Sec.~\ref{sec:3}. However, if a party does not want to send any message, he or she just needs to `vote' for the `candidate' $\bar{m-1}$.  Finally each $R_k \in\{0,1,\cdots,m-2\}$ will be the message sent by one of the senders. Therefore, each sender broadcasts the intended message anonymously.

\subsection{Anonymous ranking}
Anonymous ranking (AR) \cite{QAR} is an important problem in AMC and has significant practical applications \cite{QAR}. An AR task generally involves two steps. 1) each party needs to broadcast his or her data $y_k = \brac{y_k^0,y_k^1,...,y_k^{i_k-1}}$ to the community anonymously. 2) Each one could rank the published data by himself or herself. Obviously the first step could be done safely by using our AMC protocol. Finally instead of self-tallying like in SQAV, self-ranking is proceeded.

\section{Discussion}
We discuss in detail how our SQVA ensures privacy in Sec.~\ref{sec:4}. However, except being able to keep privacy for each voter, our protocol has several other nice properties which are not fulfilled by other existing protocols \cite{VSC,HZBB,DPT,BBHZ,HK,JHNXZ}.

1) {\em Self-tally}.
In our protocol, any voter or other third party who is interested in voting results can tally the votes by himself or herself by counting the votes in $\brac{R_j}$ in step (3.2). Then through simple calculation, they can obtain the voting result.

2) {\em Non-reusability}.
In our voting protocol, each voter cannot cast more than one votes. More specifically, each voter cannot vote one candidate more than once or vote more than one candidates. Suppose voter $V_k$ wants to vote twice $v_k$ and $v_e$ in step (3.1). To do so, he or she first casts $v_k$ to the ballot box decided by his or her index number, $d_k$ as usual, then he or she casts $v_e$ to another ballot box labeled by $d_{e}$. However since the index array $\brac{d_k}$ is a permutation of $\mathbb{Z}_n$, $d_{e}$ must be the index number of another voter $V_{j}$. Therefore $V_{j}$ will find that $R_{d_{j}} = v_{j} + v_e \neq v_{j}\mod m$ and knows that someone cheats thus aborting the voting protocol. Our protocol ensures that each voter only has one vote and he or she can only use it once.

3) {\em Verifiability}.
In our protocol step (3.3), each voter can verify if his or her vote has been modified by attackers. As long as $V_k$ finds out $R_{d_k} \neq v_k$, he or she knows that his or her vote has not been counted correctly.

4) {\em Fairness}.
 If a voter could know some useful information about other votes beforehand, he or she might change his or her mind thus voting for another candidate to his or her benefit. In our protocol, the voters vote only in the step (3) and the vote tally is obtained by doing statistics on $R_k$ which is the sum over the numbers $r_{j,k}'$. However, the numbers $r_{j,k}'$ are announced via simultaneous broad channels in the step (3.1), which means that each voter cannot acquire the other voters' information on $r_{j,k}'$ and thus cannot obtain a partial vote tally beforehand. Therefore, fairness can be maintained.

\section{Conclusion}
\label{sec:conclustion}
We have presented a quantum protocol for implementing the task of anonymous voting with the help of two entangled quantum states, $|\mathcal{X}_n\rangle$ and $|\mathcal{S}_n\rangle$. Through our protocol, any individual party can acquire a permutation of all the votes, which makes anyone can tally the votes by himself or herself without resorting to a third-party tally man. The protocol has been demonstrated to possess the properties of privacy, self-tallying, non-reusability, verifiability and fairness. We also generalize our SQAV to the more general AMC task. Our generalized protocol could let each party broadcast his or her data anonymously and safely to be further fed into AMC function.

An interesting open question is whether and how our protocol can be used to implement more tasks on AMC or SMC. This deserves further investigations in the future.

\begin{acknowledgments}
This work is supported by National Natural Science Foundation of China (Grant Nos. 61272057, 61572081) and funded by China Scholarship Council. H.Q. is supported by the Air Force Office of Scientific Research, the Army Research Office, and the National Science Foundation.
\end{acknowledgments}

\appendix
\section{Proof $\ket{\S_n}$ is n-lateral rotationally invariant}
\label{app:X}

	\noindent\textbf{Property 1} { A $n$ dimensional qudit state on Hilbert space $\mathcal{H}_{n}$ is the superposition of its computational basis $\brac{\ket{i}}~{i\in\brac{0,1,\cdots,n-1}}$ . Consider a state of $n$ such particles on $\mathcal{H}_n^{\otimes n}$,
		\begin{eqnarray}
		\ket{\S_n} &=& \sum_{S\in \mathcal{P}_n^n}(-)^{\tau(S)}\ket{S}\\\label{eq:Xexpand}
		&\equiv& \sum_{S\in \mathcal{P}_n^n}(-)^{\tau(S)}\ket{s_0s_1,\cdots,s_{n-1}}.
		\end{eqnarray}
	Consider another basis $\brac{\ket{i'}}$ connected with the computational basis by a unitary transformation $U$,
		\begin{eqnarray}\label{eq:unitary}
		\ket{i} = \sum_{j}U_{ji}\ket{j'}~.
		\end{eqnarray}
		Then in this new basis the state $\ket{\S_n}$ takes the same form up to a global phase factor $\phi$:
		\begin{eqnarray}
		\ket{\S_n} &=& e^{i\phi}\sum_{M\in \mathcal{P}_{n}^n}(-)^{\tau(M)}\ket{M'}\\
		&\equiv& e^{i\phi}\sum_{M\in \mathcal{P}_{n}^n}(-)^{\tau(M)}\ket{m_0'm_1'\cdots m_{n-1}'}~.
		\end{eqnarray}
		Here $\mathcal{P}_n^n=\{x_0 x_1\cdots x_{n-1}|x_0, x_1,\cdots, x_{n-1} \in \mathbb{Z}_n, \forall j \neq k, x_j \neq x_k\}$ and the phase factor is given by
		\begin{eqnarray}
		e^{i\phi}=\det(U)~.
		\end{eqnarray} }\\
	
	\noindent\textbf{Proof}: Expand Eq.(\ref{eq:Xexpand}) in the new basis by using the unitary transformation Eq.(\ref{eq:unitary}), we have
	\begin{eqnarray}
	\nonumber\ket{\S_n} &=& \sum_{S\in \mathcal{P}_n^n}(-)^{\tau(S)}\sum_{m_0 = 0}^{n-1}U_{m_0,s_0}\ket{m_0'}\otimes\cdots\otimes\\
	&&\sum_{m_{n-1} = 0}^{n-1}U_{m_{n-1},s_{n-1}}\ket{m_{n-1}'}\\
	\nonumber&=&(\sum_{M\in\mathcal{P}_{n}^n}+\sum_{M\notin\mathcal{P}_{n}^n})\big[\sum_{S\in \mathcal{P}_{n}^n}(-)^{\tau(S)}U_{m_0,s_0}U_{m_1,s_1}\cdots\\
	&&U_{m_{n-1},s_{n-1}}\big]\ket{M}\\
	&=& (\sum_{M\in\mathcal{P}_{n}^n}+\sum_{M\notin\mathcal{P}_{n}^n})\det(U_{m_j,s_i})\ket{M}
	\end{eqnarray}
	if $M\notin\mathcal{P}_{n}^n$, $\exists s\neq t$ such that $m_s = m_t$, then there are two same columns for matrix $U_{m_j,s_i}$, that is, $U_{m_s,s_i} = U_{m_t,s_i}$. Therefore $\det{U_{m_j,s_i}} = 0$ and we have
	\begin{eqnarray}
	\nonumber\ket{\S_n} &=& \sum_{M\in\mathcal{P}_{n}^n}\det(U_{m_j,s_i})\ket{M}\\
	\nonumber&=&\sum_{M\in\mathcal{P}_{n}^n}(-)^{\tau(M)}\det(U_{j,s_i})\ket{M}\\
	\nonumber&=& \sum_{M\in\mathcal{P}_{n}^n}(-)^{\tau(M)}\det(U)\ket{M}\\\label{}
	&=& e^{i\phi}\sum_{M\in\mathcal{P}_{n}^n}(-)^{\tau(M)}\ket{M}
	\end{eqnarray}$\blacksquare$\\
\section{Proof of Theorem 2}
\label{app:S}
To prove the theorem 2, we first give two lemmas and one corollary.\\

\noindent\textbf{Lemma 1}
Letting $q$ be an arbitrary element of  $\{1,2,\cdots,n-1\}$, $s_0,s_1,\cdots,s_{q-1} \in \mathbb{Z}_n$ are distinct, if $\sum_{j=0}^{q-1} \exp(\frac{2 \pi i s_j t}{n})\alpha_j=0$ always holds for any $t \in \mathbb{Z}_n$, we have $\alpha_0=\alpha_1=\cdots=\alpha_{q-1}=0$.\\

\noindent\textbf{Proof}:
If $\sum_{j=0}^q \exp(\frac{2 \pi i s_j t}{n})\alpha_j=0$ always holds for any $t \in \mathbb{Z}_n$, we have linear equations
 \begin{eqnarray}
 \label{EQ:A1}
A \left(
\begin{array}{c}
\alpha_0 \\
\alpha_1 \\
\vdots \\
\alpha_{q-1}
\end{array}
\right)
=\left(
\begin{array}{c}
0 \\
0 \\
\vdots \\
0
\end{array}
\right),
\end{eqnarray}
where A is a $n\times q$ matrix with elements $A_{jk}=\exp(\frac{2 \pi i (j-1)s_k}{n})=(\exp(\frac{2 \pi i s_k}{n}))^{j-1}$. Taking the first $q$ rows of $A$ as a new square matrix $\overline{A}$ with size $q \times q$ , it is easy to see that $\overline{A}$ is a Vandermonde matrix \cite{Vandermonde}. Since $s_0,s_1,\cdots,s_{q-1}$ are distinct, the determinant of $\overline{A}$ is non-zero and thus the rank of $A$ is $q$. Consequently, the above Eq. (\ref{EQ:A1}) has the only solution $\alpha_0=\alpha_1=\cdots=\alpha_{q-1}=0$.\\

\noindent\textbf{Lemma 2}
Letting $m$ be an arbitrary element of  $\{2,3,\cdots,n\}$, $\mathcal{R}_n^m=\{x_0 x_1\cdots x_{m-1}| x_0,x_1,\cdots,x_{m-1} \in \mathbb{Z}_n\}$, $\mathcal{P}_n^m=\{x_0 x_1\cdots x_{m-1}|x_0,x_1,\cdots, x_{m-1} \in \mathbb{Z}_n, \forall j \neq k, x_j \neq x_k\}$ and $\mathcal{Q}_n^m=\{x_0 x_1\cdots x_{m-1}| x_0, x_1,\cdots, x_{m-1} \in \mathbb{Z}_n, \exists j \neq k, x_j=x_k\}$. Apparently, $\mathcal{P}_n^m  \cap \mathcal{Q}_n^m = \emptyset $ and $\mathcal{R}_n^m=\mathcal{P}_n^m \cup \mathcal{Q}_n^m$. Divide $\mathcal{P}_n^m$ into ${{n}\choose{m}}=\frac{n!}{(n-m)!m!}$ subsets, each one corresponding to the set of all the $m!$ permutations of a $m$-combination of $\mathbb{Z}_n$, denoted by $\mathcal{P}_n^{n,w}~(w=0,1,\cdots,{{n}\choose{m}}-1)$. For an arbitrary subset $\mathcal{P}_n^{m,w}$, if the equation
\begin{eqnarray}
\label{EQ:A2}
\sum_{S \in \mathcal{P}_n^{m,w}} (-1)^{\tau(S)}\prod_{j=0}^{m-1}\exp(\frac{2\pi i s_j t_j}{n}) &&\beta_{S} =0
\end{eqnarray}
holds for any $t_0t_1\cdots t_{m-1} \in  \mathcal{Q}_n^m$, we have that  all the $\beta_{S}$ for $S\in \mathcal{P}_n^{m,w}$ are equal.\\

\noindent\textbf{Proof}:
We use the method of induction to prove this lemma.

For $m=2$, supposing $\mathcal{P}_n^{2,w}=\{\hat{s}_0\hat{s}_1, \hat{s}_1\hat{s}_0\}(\hat{s}_0 < \hat{s}_1)$, $\mathcal{Q}_n^2=\{t_0t_1|t_0=t_1=t \in \mathbb{Z}_n\}$ and the equation $\sum_{s_0s_1 \in \mathcal{P}_n^{2,w}} (-1)^{\tau(s_0s_1)}\exp(\frac{2\pi i (s_0 t_0+s_1t_1)}{n}) \beta_{s_0s_1}=0$ holds for any $t_0t_1 \in  \mathcal{Q}_n^2$. Since $t_0=t_1=t$, the equation can also be written as $\exp(\frac{2\pi i(\hat{s}_0+\hat{s}_1)t}{n}) \beta_{\hat{s}_0\hat{s}_1}-\exp(\frac{2\pi
i(\hat{s}_0+\hat{s}_1)t}{n}) \beta_{\hat{s}_1\hat{s}_0}=0$. Obviously, $\beta_{\hat{s}_0\hat{s}_1}=\beta_{\hat{s}_1\hat{s}_0}$ is obtained.

We assume that for $m=k$ and an arbitrary subset $\mathcal{P}_n^{k,w}$, if the Eq. (\ref{EQ:A2}) always holds for any $t_0t_1\cdots t_{k-1} \in  \mathcal{Q}_n^k$, all the $\beta_{s_0s_1\cdots s_{k-1}}$ for $s_0s_1\cdots s_{k-1} \in \mathcal{P}_n^{k,w}$ are equal. Now we analyze the case for $m=k+1$. We suppose the ($k$+1)-combination that $\mathcal{P}_n^{k+1,w}$ corresponds is the set $\hat{S}=\{\hat{s}_0,\hat{s}_1,\cdots,\hat{s}_{k}\}$ with $\hat{s}_0<\hat{s}_1<\cdots<\hat{s}_{k}$, namely, $\mathcal{P}_n^{k+1,w}$ is the set of all the $(k+1)!$ permutation of the $\hat{S}$. In this case, observing that $s_p$ ($p \in \{0,1,\cdots,k\}$) can take all the values from $\hat{S}$ in the Eq. (\ref{EQ:A2}), the equation can be written as

\begin{eqnarray}
\label{EQ:A3}
\sum_{l=0}^{k}
\Bigg(
&&\sum_{S \in \mathcal{P}_n^{k+1,w}, s_p=\hat{s}_l} (-1)^{\tau(S)}\exp(\frac{2\pi i \hat{s}_l t_p}{n}) \nonumber \\
&& \prod_{j=0,j\neq p}^{k}\exp(\frac{2\pi i s_j t_j}{n}) \beta_{S}
\Bigg)=0.
\end{eqnarray}
Noting that $(-1)^{\tau(S)}=(-1)^{l-p}(-1)^{\tau(s_0\cdots s_{p-1} s_{p+1}\cdots s_{k})}$, the Eq. (\ref{EQ:A3}) can also be written as

\begin{widetext}
\begin{eqnarray}
\label{EQ:A4}
\sum_{l=0}^{k}(-1)^{l-p}\exp(\frac{2\pi i \hat{s}_l t_p}{n})
\Bigg(
\sum_{S \in \mathcal{P}_n^{k+1,w}, s_p=\hat{s}_l}
(-1)^{\tau(s_0 \cdots s_{p-1} s_{p+1}\cdots s_{k})}\prod_{j=0,j\neq p}^{k}\exp(\frac{2\pi i s_j t_j}{n})\beta_{S}
\Bigg)=0.&& \nonumber \\
 &&
\end{eqnarray}
\end{widetext}
We now prove that, if the Eq. (\ref{EQ:A4}) holds for any $t_0t_1\cdots t_{k} \in  \mathcal{Q}_n^{k+1}$, all the $\beta_{S}$ for $S \in \mathcal{P}_n^{k+1,w}$ are equal. Specially, when $t_0\cdots t_{p-1}t_{p+1}\cdots t_{k} \in \mathcal{Q}_n^{k}$ is fixed and $t_p$ takes every value from $\mathbb{Z}_n$, the Eq. (\ref{EQ:A4}) always holds. Hence, according to the lemma 1, we can derive that for arbitrary $l \in \{0,1,2,\cdots,k\}$,
\begin{eqnarray}
\label{EQ:A5}
&&\sum_{S \in \mathcal{P}_n^{k+1,w}, s_p=\hat{s}_l}(-1)^{\tau(s_0\cdots s_{p-1}s_{p+1} \cdots s_{k})} \nonumber \\
&&\prod_{j=0,j\neq p}^{k}\exp(\frac{2\pi i s_j t_j}{n}) \beta_{s_0 s_1 \cdots s_{k}}=0.
\end{eqnarray}
Meanwhile, since the Eq. (\ref{EQ:A5}) holds for arbitrary $t_0\cdots t_{p-1}t_{p+1}\cdots t_{k} \in \mathcal{P}_n^{k,w}$, based on the previous assumption for the case $m=k$, all the $\beta_{S}$ for $S \in \mathcal{P}_n^{k+1,w}$ and $s_p=\hat{s}_l$ are equal. Since $l$ and $p$ can take arbitrary values from $\{0,1,2,\cdots,k\}$, we can draw the conclusion that for $m=k+1$, if the equation (\ref{EQ:A2}) holds for any $t_0t_1\cdots t_{k} \in  \mathcal{Q}_n^{k+1}$, we have that  all the $\beta_{S}$ for $S \in \mathcal{P}_n^{k+1,w}$ are equal.

By mathematical induction above, we can derive that for arbitrary $m \in \{2,\cdots,n\}$, if the Eq. (\ref{EQ:A2}) holds for any $t_0t_1\cdots t_{m-1} \in  \mathcal{Q}_n^m$,  all the $\beta_{s_0s_1\cdots s_{m-1}}$ for $s_0s_1\cdots s_{m-1} \in \mathcal{P}_n^{m,w}$ are equal.\\

Now we give a corollary of lemma 2 below.\\

\noindent\textbf{Corollary 1}
Let $m$, $\mathcal{R}_n^m$, $\mathcal{P}_n^m$, $\mathcal{Q}_n^m$ and $\mathcal{R}_n^m$ be defined as lemma 2. For an arbitrary subset $\mathcal{P}_n^{m,w}$, if the equation
\begin{eqnarray}
\label{EQ:A6}
\sum_{s_0s_1\cdots s_{m-1} \in \mathcal{P}_n^{m,w}} (-1)^{\tau(s_0s_1\cdots s_{m-1})}\prod_{j=0}^{m-1}\exp(\frac{2\pi i s_j t_j}{n}) && \vec{\beta}_{s_0s_1\cdots s_{m-1}} \nonumber \\
&& = \vec{0}
\end{eqnarray}
holds for any $t_0t_1\cdots t_{m-1} \in  \mathcal{Q}_n^m$, where $\vec{\beta}_{s_0s_1\cdots s_{m-1}}$ are vectors and $\vec{0}$ is zero vector, all the $\vec{\beta}_{s_0s_1\cdots s_{m-1}}$ for $s_0s_1\cdots s_{m-1} \in \mathcal{P}_n^{m,w}$ are equal.

The only difference between this corollary and lemma 2 is that $\beta_{s_0s_1\cdots s_{m-1}}$ is generalized to the vector $\vec{\beta}_{s_0s_1\cdots s_{m-1}}$.  Hence, the corollary can be directly proved.\\

Now we use lemma 2 to prove theorem 2.\\

\noindent\textbf{Proof}:
Restrict the measurement basis to computation basis or fourier basis, the necessity of our theorem can be directly obtained from the property 1.

Now we prove the sufficiency. On one hand, to satisfy the condition that all the measurement outcomes are distinct when measuring each particle of $|\Theta\rangle$ in fourier basis, $|\Theta\rangle$ must be in the form
\begin{widetext}
\begin{eqnarray}
\label{EQ:A7}
|\Theta\rangle &=& \sum_{S \in \mathcal{P}_n^n} (-1)^{\tau(S)} \beta_S(\mathcal{F}|s_0\rangle)\otimes \cdots \otimes(\mathcal{F}|s_{n-1}\rangle) \nonumber \\
&=&\sum_{S \in \mathcal{P}_n^n}(-1)^{\tau(S)}\beta_S(\sum_{t_0}\frac{\exp(\frac{2\pi i s_0t_0}{n})}{\sqrt{n}}|t_0\rangle)\otimes \cdots \otimes(\sum_{t_{n-1}}\frac{\exp(\frac{2\pi i s_{n-1}t_{n-1}}{n})}{\sqrt{n}}|t_{n-1}\rangle)\nonumber \\
&=&\sum_{t_0,t_1,\cdots,t_{n-1}} \sum_{S \in \mathcal{P}_n^n} (\frac{(-1)^{\tau(S)}}{n^{\frac{n}{2}}} \prod_{j=0}^{n-1}\exp(\frac{2 \pi i s_j t_j}{n})\beta_S)|t_0 t_1 \cdots t_{n-1}\rangle,
\end{eqnarray}
\end{widetext}
where $S=s_0s_1\cdots s_{n-1}$. On the other hand, to meet the condition that all the measurement outcomes are distinct when measuring each particle of $|\Theta\rangle$ in computation basis,
the terms $\sum_{S \in \mathcal{P}_n^n} (\beta_S \prod_{j=0}^{n-1}\exp(\frac{2 \pi i s_j t_j}{n}))$ for $t_0t_1\cdots t_{n-1} \in \mathcal{Q}_n^n$ are required to be equal to zero. From lemma 2 (when $m=n$), to satisfy this requirement, we can see that all the $\beta_S$ for $S \in \mathcal{P}_n^n$ are equal. Moreover, to keep normalization of $|\Theta\rangle$, we have
\begin{eqnarray}
\label{EQ:A8}
\beta_S=\frac{1}{\sqrt{n!}}.
\end{eqnarray}

For any $t_0t_1\cdots t_{n-1} \in \mathcal{Q}_n^n$, according to the definition of square matrix determinant, $\sum_{S \in \mathcal{P}_n^n} \frac{(-1)^{\tau(S)}}{n^{\frac{n}{2}}} \prod_{j=0}^{n-1}\exp(\frac{2 \pi i s_j t_j}{n})$ is in fact the determinant of the $n\times n$ matrix $\overline{V}$ with elements $\overline{V}_{jk}=\frac{\exp(\frac{2\pi i t_j k}{n})}{\sqrt{n}}$, namely,
\begin{eqnarray}
 \label{EQ:A9}
\sum_{S \in \mathcal{P}_n^n} \frac{(-1)^{\tau(S)}}{n^{\frac{n}{2}}} \prod_{j=0}^{n-1}\exp(\frac{2 \pi i s_j t_j}{n})= \det(\overline{V}).
\end{eqnarray}
Transposing pairs of rows of $\overline{V}$ to generate a new $n\times n$ matrix $\widetilde{V}$ with elements $\widetilde{V}_{jk}=\frac{\exp(\frac{2\pi i jk}{n})}{\sqrt{n}}$, we have
\begin{eqnarray}
 \label{EQ:A10}
\det(\overline{V})=(-1)^{\tau(t_0t_1\cdots t_{n-1})}\det(\widetilde{V}).
\end{eqnarray}
Taking the Eqs. (\ref{EQ:A8}), (\ref{EQ:A9}) and (\ref{EQ:A10}) to the Eq. (\ref{EQ:A7}) and discarding the terms for $t_0t_1\cdots t_{n-1} \in \mathcal{Q}_n^n$ in the Eq. (\ref{EQ:A7}), we have
\begin{eqnarray}
 \label{EQ:A12}
|\Theta\rangle = \sum_{T \in \mathcal{P}_n^n}\frac{(-1)^{\tau(T)}}{\sqrt{n!}}|T\rangle,
\end{eqnarray}
up to the global factor $\det(\widetilde{V})$, where $T=t_0t_1\cdots t_{n-1}$. Therefore, $|\Theta\rangle$ has the same form as $|\mathcal{S}_n\rangle$ and the theorem 2 is proved.

\bibliography{basename of .bib file}
\bibliography{apssamp}

\end{document}